\documentclass[aps,prb,twocolumn,showpacs,floatfix]{revtex4}

\usepackage{multirow}
\usepackage{graphicx}
\usepackage{amsmath}
\usepackage{subfigure}

\def\beq{\begin{eqnarray}}
\def\eeq{\end{eqnarray}}

\begin{document}

\title{Comment on "Bound states of edge dislocations: The quantum dipole problem in two dimensions"}
\author{Paolo Amore}
\email{paolo.amore@gmail.com}
\affiliation{Facultad de Ciencias, CUICBAS,Universidad de Colima, \\
Bernal D\'{\i}az del Castillo 340, Colima, Colima, Mexico} 
\date{\today}

\begin{abstract}    
We show that it is possible to improve some of the numerical results contained in a recent paper 
with an optimal implementation of the methods used in that paper. A careful analysis done using 
the Rayleigh-Ritz method provides a rigorous upper bound for the energy of the ground state of an 
electron in a two dimensional potential generated by the edge dislocation, as well as precise 
values for the excited states. The extrapolation of the results corresponding to different subspaces 
is used to obtain an alternative estimate of the fundamental energy of the model. 
The energies of the first 500 states that we have calculated are in perfect agreement
with the expected asymptotic behavior.
\end{abstract}
\maketitle

The presence of edge dislocations in a solid may have strong effects on the mechanical, transport, elastic and 
superconducting properties of the solid~\cite{Dorsey10}. The knowledge of the spectrum of the bound states at the edge is therefore
helpful to assess the change in the properties of the material. 

With this motivation, the authors of a recent paper, ref.~\cite{Dorsey10}, have considered the Schr\"odinger equation (SE)
for an electron confined in a two dimensions and subject to a potential  $V(r,\theta) =  p \cos\theta /r$
(corresponding to a straight dislocation along the $z$-axis):
\beq
- \frac{\hbar^2}{2m} \nabla^2 \Psi + p \frac{\cos \theta}{r} \Psi = E \Psi \ .
\label{eq_SE_dipole}
\eeq

Unfortunately this equation is not exactly solvable and its solution requires approximate methods. 
In effect ref.~\cite{Dorsey10} provides a detailed list of the different approaches which have been applied in the literature 
to this problem~\cite{Landauer54,Emtage67,Shapiro77,Chishko84,Dubrovskii97,Farvacque01}: the numerical estimates
for the energy of the ground state obtained in these works are reported in Table 1 of ref.\cite{Dorsey10} in units of $2mp^2/\hbar^2$.
The last result of the table is the improved estimate obtained in that paper discretizing the Schr\"odinger equation on a uniform square
grid and working with matrices of a maximum size of $10^6 \times 10^6$. The authors quote an error of $2 \%$ for the numerical eigenvalues
of the planar Coulomb problem, whose exact solutions are known~\cite{Yang91}. Unfortunately, since the precision of the results depends on
the specific problem considered, the accuracy of the numerical results of ref.~\cite{Dorsey10} is not granted for the dipole problem 
and actually one may expect on qualitative grounds  larger errors for the quantum dipole problem, due to the dependence on $\theta$ in the potential.

In both cases, the long range nature of the potential and its singular behavior
at $r=0$ would better be taken into account using a nonuniform two-dimensional grid (see refs.~\cite{Fattal96, Boyd00}
for a discussion of collocation methods with nonuniform grids).
 
A second difficulty of the computational approach based on the discretization of the Schr\"odinger equation, which is mentioned in 
ref.~\cite{Dorsey10}, is the limited number of states which can be obtained with acceptable precision with a given grid, due to the 
different length scales of the excited states: as a result the estimates for the first few states of the dipole potential cannot be 
equally accurate. Finally, one should bear in mind that the discretization of the SE, however accurate it could be, does not 
provide  upper bounds on the energy of the fundamental mode: the result obtained in ref.~\cite{Dorsey10} may thus either over or
underestimate the exact energy.

The second approach discussed by the authors of ref.\cite{Dorsey10} is what they call a "Coulomb basis method", which is essentially 
a Rayleigh-Ritz (RR) approach which uses the basis of the planar Coulomb problem (see ref.~\cite{Yang91}). The Rayleigh-Ritz approach,
in contrast to the real-space diagonalization method mentioned earlier, does provide an upper bound to the energy of the fundamental state
and a direct decomposition of the approximate solutions in the basis chosen. Unfortunately the potentialities of this method have not been 
fully exploited in that paper: the main purpose of this Comment is then to obtain stricter bounds for the energy of the ground state working with
a larger set of functions. 

We proceed to describe the approach; the 2d hydrogen wave functions are~\cite{Yang91}
\beq
\psi_{nl}(r,\theta) &=& R_{nl}(r) \ \chi_l(\theta) 
\label{basid_2d_hyd}
\eeq
with $n=1,2, \dots$ and $-n+1 \leq l \leq n-1$.

The angular and radial parts  respectively read~\footnote{Notice the typo in the bounds over $l$ in eq.(5) of 
ref.~\cite{Dorsey10}.}
\beq
\chi_l(\theta) &\equiv& \frac{1}{\sqrt{\pi}} \left\{ \begin{array}{ccc}
\cos l\theta & , & 1\leq l \leq n-1 \\ 
1/\sqrt{2}   & , & l=0 \\
\sin l\theta & , & -n+1\leq l \leq -1 \\ 
\end{array}
\right.
\eeq
and
\beq
R_{nl}(r) &\equiv& \frac{\beta_n}{2|l|!} \sqrt{\frac{(n+|l|-1)!}{(2n-1) (n-|l|-1)!}} \ \left(\beta_n r \right)^{|l|} \nonumber \\
&\cdot& e^{-\beta_n r/2} \ _1F_1\left(-n+|l|+1,2|l|+1,\beta_n r\right) \ ,
\eeq
where $\beta_n \equiv \frac{4}{(2n-1)} \frac{m e^2}{\hbar^2}$. Here $_1F_1(a,b,c)$ is the confluent hypergeometric function.

The bound state energies of the 2D hydrogen are simply~\cite{Yang91}
\beq
\epsilon_{n} = - \frac{2 m e^4}{\hbar^2 (2 n-1)^2} \ .
\eeq

The RR approach requires the calculation of the matrix elements of the Hamiltonian of eq.~(\ref{eq_SE_dipole})
in the basis (\ref{basid_2d_hyd}):
\beq
\langle n_1 l_1 | && \hskip-0.26in \hat{H} | n_2 l_2 \rangle = 
\delta_{l_1 l_2} \delta_{n_1,n_2} \epsilon_{n_1} \nonumber \\
&+& 
\left( e^2 \mathcal{I}^{(1)}_{l_1,l_2} + p \mathcal{I}^{(2)}_{l_1,l_2} \right) 
\left[ \int_0^\infty R_{n_1l_1}(r) R_{n_2l_2}(r) dr\right] \nonumber \ ,
\eeq
where
\beq
&& \mathcal{I}^{(1)}_{l_1,l_2} \equiv \int_0^{2\pi}  \chi_{l_1}(\theta) \chi_{l_2}(\theta) d\theta = \delta_{l_1,l_2} \nonumber \\
&& \mathcal{I}^{(2)}_{l_1,l_2} \equiv  \int_0^{2\pi}  \chi_{l_1}(\theta) \cos\theta \chi_{l_2}(\theta) d\theta 
= \left[\delta_{l_1+1,l_2}+\delta_{l_1-1,l_2} \right] \nonumber \\
&\cdot&\left\{ \frac{\theta(l_1)\theta(l_2) + \theta(-l_1)\theta(-l_2)}{2} 
+ \frac{\delta_{l_1,0}\theta(l_2)  + \delta_{l_2,0}\theta(l_1)}{\sqrt{2}} 
\right\} \nonumber \ .
\eeq

The evaluation of the radial integral is not straightforward as in the case of the angular integrals but it can also be done
analytically. Defining $a_i \equiv -n_i+|l_i|+1$ and $b_i\equiv 2|l_i|+1$ ( $i=1,2$), the confluent hypergeometric functions in this 
integral reduce to polynomials of degrees $-a_1$ and $-a_2$ respectively, for $a_{1,2}< 0$ and $b_{1,2}$ positive integers.
The original integral is therefore reduced to a sum of integrals which can be done explicitly: 
\beq
\int_0^\infty R_{n_1l_1}(r) R_{n_2l_2}(r) dr &=&  
\frac{m e^2}{\hbar^2} \ \sum_{j_1=0}^{|a_1|} \sum_{j_2=0}^{|a_2|}  \frac{\mathcal{N}_{n_1,l_1,j_1;n_2,l_2,j_2}}{\mathcal{D}_{n_1,l_1,j_1;n_2,l_2,j_2}} \ ,
\eeq
where the $(a)_k$ are the Pochhammer symbols and
\beq
\mathcal{N}_{n_1,l_1,j_1;n_2,l_2,j_2} &\equiv& 4  (2 n_2-1)^{|l_1|+j_1-\frac{1}{2}} (2n_1-1)^{|l_2|+j_2-\frac{1}{2}} \nonumber \\
&& (-n_1+|l_1|+1)_{j_1} 
(-n_2+|l_2|+1)_{j_2} \nonumber \\
&\cdot& \Gamma(j_1+j_2+|l_1|+|l_2|+1) \nonumber \\
&& \sqrt{\Gamma (n_1+|l_1|) \Gamma (n_2+|l_2|)} \nonumber \\
&& (n_1+n_2-1)^{-|l_1|-|l_2|-j_1-j_2-1} \\
\mathcal{D}_{n_1,l_1,j_1;n_2,l_2,j_2} &\equiv& \Gamma (j_1+1) \Gamma (j_2+1) \nonumber \\
&&\Gamma (j_1+2 |l_1|+1) \Gamma(j_2+2 |l_2|+1) \nonumber \\
&& \sqrt{\Gamma (n_1-|l_1|) \Gamma (n_2-|l_2|)} \ .
\eeq

The authors of ref.\cite{Dorsey10} have used the analytical expressions for these integrals, although they had to 
restrict their calculation to a set of $400$ basis functions, corresponding to $-n+1 \leq l \leq n-1$ and  $1 \leq n \leq 20$, 
due to the numerical round-off errors that become important for the matrix elements corresponding to larger quantum numbers. 
In our numerical calculation we have used Mathematica 8 ~\cite{wolfram}, obtaining symbolic expressions for the matrix elements,
which were then evaluated numerically avoiding the round-off errors which would appear in a fully numerical calculation.

It is convenient to introduce the notation $\alpha  \equiv e^2$ and regard $\alpha$ as a variational parameter
controlling the length scale.  We then rewrite the matrix elements of the Hamiltonian making the dependence upon $\alpha$ explicit:
\beq
&& \langle n_1 l_1 | \hat{H} | n_2 l_2 \rangle 
= \frac{m \alpha^2}{\hbar^2} \delta_{l_1 l_2} \left[ - \delta_{n_1,n_2} \frac{2}{(2 n-1)^2} \right. \nonumber \\ 
&+& \left. \sum_{j_1=0}^{|a_1|} \sum_{j_2=0}^{|a_2|}  \frac{\mathcal{N}_{n_1,l_1,j_1;n_2,l_2,j_2}}{\mathcal{D}_{n_1,l_1,j_1;n_2,l_2,j_2}} \right]
+  \frac{\alpha m p}{\hbar^2} \ \mathcal{I}^{(2)}_{l_1,l_2} \nonumber \\
&\cdot& \sum_{j_1=0}^{|a_1|} \sum_{j_2=0}^{|a_2|}  \frac{\mathcal{N}_{n_1,l_1,j_1;n_2,l_2,j_2}}{\mathcal{D}_{n_1,l_1,j_1;n_2,l_2,j_2}} \ .
\eeq

This is precisely the approach followed by Dasbiswas et al. in ref.~\cite{Dorsey10}, who observe that the bound for the ground 
state energy obtained for $\alpha=1$ is not good: 
after minimizing with respect to $\alpha$ they obtain an improved bound $E_1 \geq -0.1257$ working with $400$ states (we adopt their
convention of reporting the energies in units of $2mp^2/\hbar^2$). 

Interestingly, these authors use the original basis ($\alpha=1$) for the excited states, claiming that {\sl "the real-space diagonalization 
methods provide a better estimate of the ground-state energy whereas the Coulomb basis method is more suitable for higher excited states."} 
It is not clear on what grounds this statement is made and actually we will show in this Comment that  the choice of calculating the excited 
states for $\alpha=1$ is far from optimal. We will also obtain an alternative estimate for the energy of the ground state of the 
quantum dipole problem, which falls above the one calculated in ref.~\cite{Dorsey10}. 

Our first observation concerns the number of bound states which are obtained in the calculation at a given $\alpha$: while for $\alpha = 1$ 
there are $149$ bound states, for $\alpha_{var} = 89.57$ (corresponding to the minimum of $E_1$) there are just 3 bounds states. This behavior should 
not surprise us, since $\alpha$ determines the radial length scale of the wave function, and different states have different range: 
the fact that for a large $\alpha$ fewer bound states are 
present, simply tells us that the physical states which are not captured by the calculation have a larger range. Therefore, if one wants to 
estimate a few excited states, one needs to calculate these states using different values of $\alpha$ to account for the different
length scales of each state.

The fundamental  question is therefore how to choose $\alpha$: if the problem under consideration has certain symmetries, which are also 
symmetries of the basis, then the variational principle applied to a trial wave function with that symmetry will provide again an upper 
bound to the lowest mode in that symmetry class: for our problem this is the case of functions which are odd with respect to the change 
$y \rightarrow -y$. For the remaining states, the eigenvalues of the RR matrix will vary with $\alpha$, without providing a variational 
bound. However, if we consider a given state, its exact energy and wave function will be independent of $\alpha$, which is an unphysical 
parameter of the  basis. As such, we may argue that the optimal value of $\alpha$ (and correspondingly the most accurate value 
for the energy) will be the one for which the calculated eigenvalue is less sensitive to changes of $\alpha$. This is the essence of the 
"principle of minimal sensitivity" (PMS)~\cite{Ste81}. We will soon use the PMS to calculate the excited states of eq.~(\ref{eq_SE_dipole}).

We now proceed to illustrate our numerical results, concentrating for the moment on the ground state.

\begin{figure}[t]
  \begin{center}
    \subfigure[Variational bound for the ground state energy]{\label{fig:a}\includegraphics[width = 2.5 in]{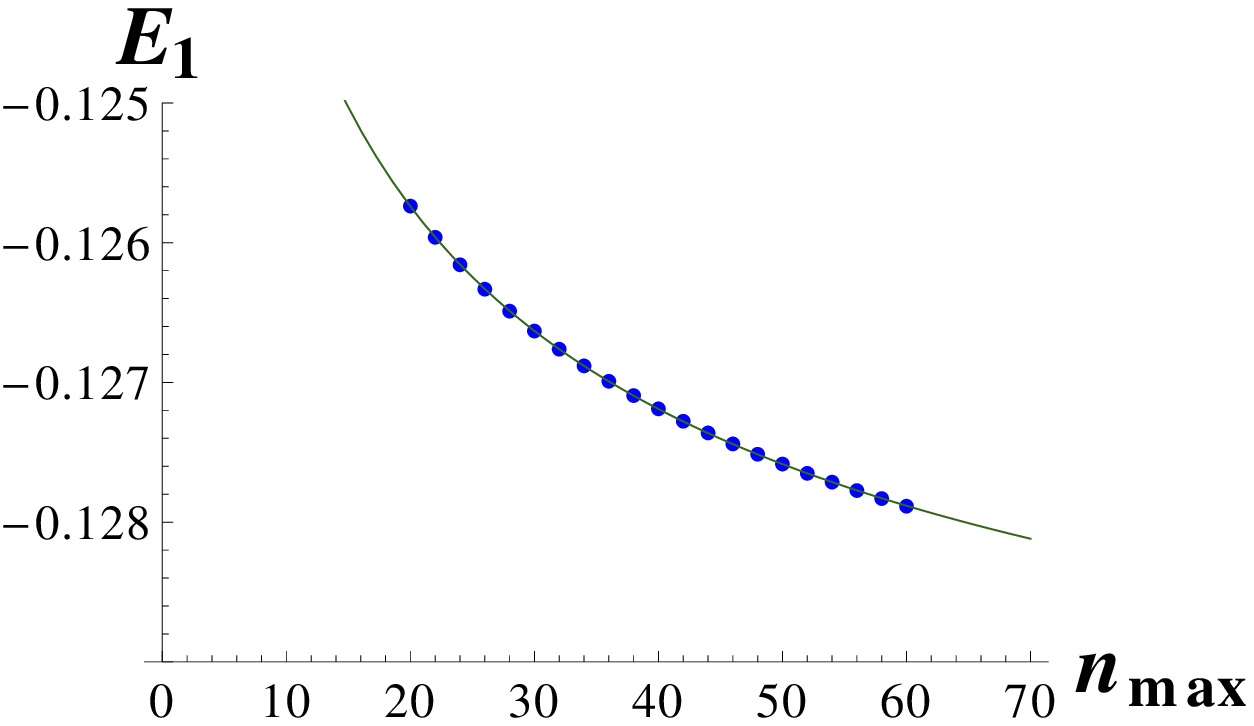}}
    \subfigure[Extrapolation of the ground state energy for $n_{max} \rightarrow \infty$]{\label{fig:b}\includegraphics[width = 2.5 in]{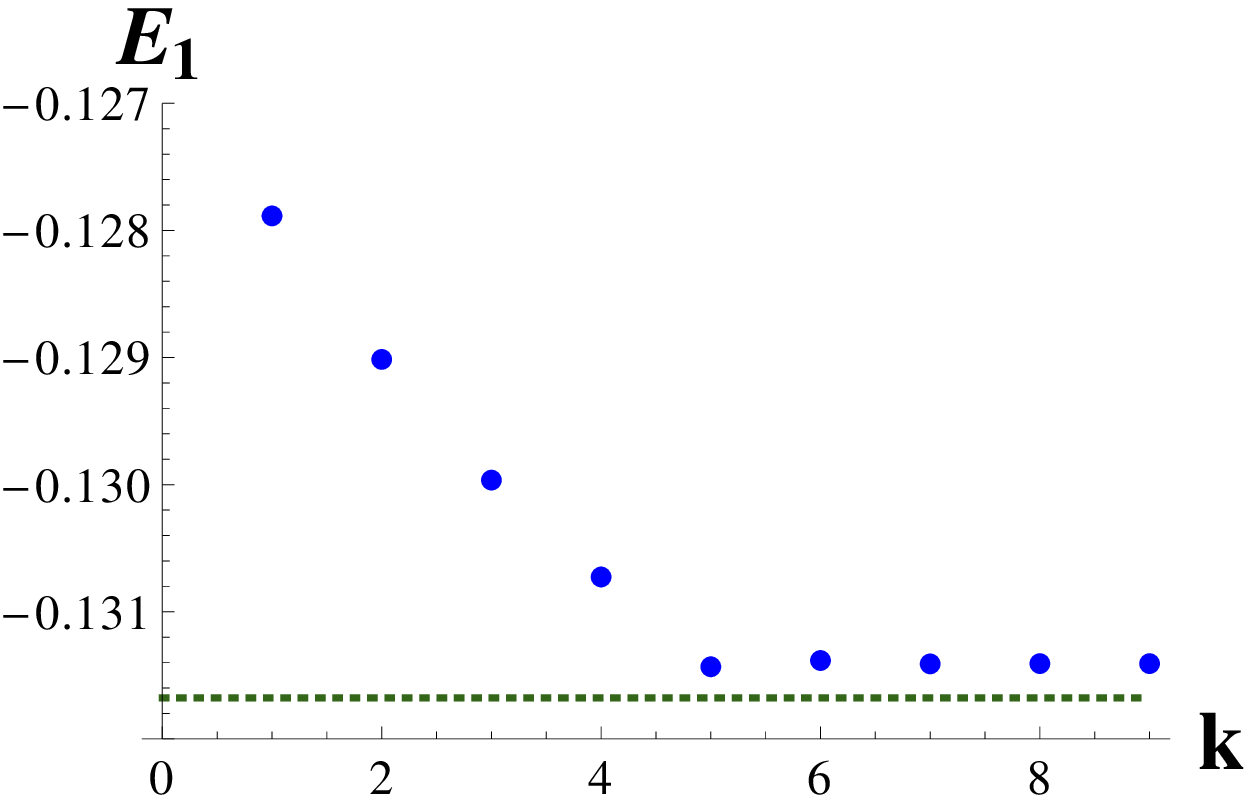}} \\
  \end{center}
  \caption{(Color online) Variational bound for $E_1$ as function of the maximum quantum number used in the RR approach (a) 
and the extrapolation of $E_1$ using the Shanks transformation; the
horizontal line corresponds to the fit of eq.~(\ref{fit2}), (b).}
 \label{Fig_1}
\end{figure}

In the first plot of Fig.\ref{Fig_1} we show the variational bound for the ground state energy of eq.(\ref{eq_SE_dipole}) as a function of the 
maximum quantum number used in the RR method (we have considered even values of $n_{max}$). For a given $n_{max}$, the portion of Hilbert space used in the calculation contains $n_{max}^2$ 
states (the results obtained by Dasbiswas et al., for instance, correspond to $n_{max}=20$ and therefore to a $400 \times 400$ matrix). 
For each value of $n_{max}$ we have selected the value of $\alpha$  for which the corresponding $E_1$ is minimal; the numerical calculation
is stopped when the convergence on the first $20$ digits of $E_1$ was reached. 

Remarkably these data display a very regular behavior, which is well described by the fit
\beq
E_1^{({\rm FIT})} = -0.131678 +\frac{0.0184}{(0.661 n+1)^{0.426}} \ .
\label{fit2}
\eeq

In the second plot of Fig.\ref{Fig_1} we show the values for the ground state energy obtained repeatedly using the Shanks transformation on the sets
of the energies of the first plot. The index $k$ in the horizontal axis indicates to the number of consecutive 
Shanks transformations used, while the value on the vertical axis is the corresponding result obtained with the Shanks transformations, 
involving the values corresponding to largest $n_{max}$. These values seem to converge to the value $E_1 \approx -0.1314$. 
The horizontal line corresponds to the constant value $E_1 = -0.131678$ reached by eq.~(\ref{fit2}) for $n_{max}\rightarrow \infty$.
Notice the disagreement of our results with the value obtained in ref.~\cite{Dorsey10}, $E_1 = -0.139$. Being fair, there is no rigorous criterion
granting that our result is more precise, although the extrapolation of very regular sequences of numbers typically provides very accurate estimates.

The behavior of the optimal $\alpha$ obtained with the PMS at different $n_{max}$ is described  very well by the 
a cubic fit $\alpha^{({\rm FIT})} = -0.00003847 n_{max}^3+0.2104 n_{max}^2+0.3062 n_{max}-0.4347$.
There is a clear physical justification of the behavior of $\alpha$, which grows with $n_{max}$: as the number of states
in the calculation is increased, one can use a basis with shorter length scale (i.e. larger $\alpha$) to build the approximate
eigenfunctions of the problem. 

Although we believe that the values of $E_1$ that we have obtained with the fit (\ref{fit2}) or with the Shanks transformation are more precise than the
result obtained in ref.~\cite{Dorsey10}, we are aware that the extrapolation of the values of $E_1$ do not themselves fulfill a variational
bound. In other words,  we may expect them to be closer to the exact value, but we cannot be sure that they fall above it.
We will now show that it is possible to obtain stricter bounds on $E_1$, even working with less states. 
As we have mentioned before, the largest set of states used in Fig.\ref{Fig_1} corresponds to  $n_{max}=60$, i.e. $3600$ states 
($1 \leq n \leq 60$ and $-n+1 \leq l \leq n-1$).  We may wonder to what extent the results would change by restricting the states to $0 \leq l \leq l_{max}$.
While one can safely drop the negative values of $l$, since the ground state must be symmetric with respect to $y \rightarrow -y$, it
is not a priori clear the error introduced by using an upper cutoff $l_{max}$.

Let $V = (v_1,v_2, \dots, v_{3600})$ be the eigenvector corresponding to the smallest eigenvalue 
of the $3600 \times 3600$ matrix obtained in RR approach: the normalization of $V$ implies that $\sum_{j=1}^{3600} v_j^2 = 1$. 
Any deviation from 1 of the sum when the components of $V$ corresponding to $l > l_{max}$ are set to zero will give us an idea of how important
these states are for the calculation. For $l_{max}=5$ one finds out that this deviation is completely negligible, $\delta \equiv 2.2 \times 10^{-9}$.
To confirm this finding we may compare the lowest eigenvalue of the full $3600 \times 3600$ matrix, $E_1 < E_1^{({\rm full})} = -0.12788587515854182342$, with 
the lowest eigenvalue of the reduced $345 \times 345$ matrix, corresponding to $l_{max} = 5$, $E_1 < E_1^{({\rm reduced})} = -0.12788587377192016287$.
We have $E_1^{({\rm reduced})} -E_1^{({\rm full})} = 1.4 \times 10^{-9}$, which is of the same order of the deviation discussed above.
The effective dimensionless coupling constant calculated with the wave function corresponding to $E_1^{({\rm reduced})}$ is $g = 0.017$, which agrees with
the result found in ref.\cite{Dorsey10}, using a simple variational ansatz.

With this result in mind we have built  a $585 \times 585$ matrix, corresponding to $n_{max}= 100$ and $l_{max}=5$, obtaining the bound $E_1 < -0.12864468596110909173$.
Notice that the fit of eq.~(\ref{fit2}) for $n_{max}=100$ provides a result which is very close to this,  $E_1^{({\rm fit})} = -0.128612$.
We have then built a $885 \times 885$ matrix, corresponding to $n_{max}= 150$ and $l_{max}=5$, obtaining our most precise bound $E_1 < -0.1291697936750557573$.
Also in this case the fit of eq.~(\ref{fit2}) for $n_{max}=150$ provides a result which is very close to this,  $E_1^{({\rm fit})} = -0.129093$.

We now discuss the excited states of eq.~(\ref{eq_SE_dipole}): as we have mentioned earlier there is no valid reason for calculating 
the excited states at $\alpha=1$. Fig.4 should convince the reader of this point: here the energy of the fifth state is plotted at different values 
of $\alpha$, and compared with the value at $\alpha =1$ (dashed line), which corresponds to the choice done in \cite{Dorsey10}. The 
dotted line at the bottom corresponds to the minimum of the curve, and it provides the most accurate value for $E_5$ which can be obtained
working with a subspace corresponding to $n_{max}=60$. Table \ref{tab1} illustrates the different results obtained using the two approaches
for the first five states. The last column reports the quantity  $\Delta_n \equiv (E_n^{(PMS)} - E_n^{(\alpha = 1)})/(E_n^{(PMS)} + E_n^{(\alpha = 1)})$,
which provides an estimate of the error done using $\alpha = 1$. Interestingly the values obtained with the PMS for the third and fifth 
states are close to the ones obtained in ref.~\cite{Dorsey10} using a discretization of the Schr\"odinger equation: these states 
correspond to smaller values of $\alpha_{PMS}$, indicating that their length scales are larger than those of the other three states.
So, for example, the third state, has a smaller $\alpha$ than the fourth state, which is higher in energy. 
The authors of \cite{Dorsey10} have also observed a similar behavior for some of the states ($23^{th}$ and $24^{th}$) that they have calculated,
although they have not given "any satisfactory explanation  for these irregular features". Our explanation of this 
phenomenon is simple: if a state with a modest spatial extent has a larger probability density in the region close to $x = 0$ (recall that the potential 
is attractive for $x <0$), its energy may be higher than the energy of a state with larger spatial extent but smaller probability density in the region 
close to $x=0$.

In Fig.~\ref{Fig_5} we plot the optimal value of $\alpha$ as a function of the level number $n$, obtained using the subspace
corresponding to $n_{max}=60$. The assumption used in ref.\cite{Dorsey10}, $\alpha = 1$, is clearly valid only for the higher 
excited states ($n > 100$). Notice the oscillations of $\alpha_{PMS}$, which signal the presence of contiguous states with different length scales,
as mentioned earlier.

In Fig.~\ref{Fig_6} we have compared the behavior of $-1/E_n$ obtained using either the PMS (solid line) or setting $\alpha =1$ (dashed line), with
the asymptotic law of eq.(10) of ref.\cite{Dorsey10} (dotted line). The plot clearly proves the superiority of the PMS results.
This superiority can also be established by looking at the fit of the first 500 values of  $-1/E_n$:
\beq
\left. -1/E_n \right|_{\alpha=1}   &=& 16.0528 n+1.384 \sqrt{n} - 2.085 \nonumber , \\
\left.-1/E_n\right|_{\alpha_{PMS}} &=& 15.9897 n+0.431 \sqrt{n} - 9.162 \nonumber \ .
\eeq

\begin{table}
\caption{\label{tab1} Comparison of the energies of the first five states of the "dipole" potential calculated with $\alpha=1$ (second column)
and using the PMS (third column) for $n_{max}=60$. Here $\Delta_n \equiv (E_n^{(PMS)} - E_n^{(\alpha = 1)})/(E_n^{(PMS)} + E_n^{(\alpha = 1)})$.
}
\begin{ruledtabular}
\begin{tabular}{ccccc}
$n$ & $E_n^{(\alpha = 1)}$ & $E_n^{(PMS)}$ & $\alpha_{PMS}$ & $\Delta_n$ \\
\hline
1 & -0.0970117 & -0.127886 & 767.132 & 0.137281 \\
2 & -0.0328379 & -0.0394579 & 294.189 & 0.0915679\\
3 & -0.0220914 & -0.0232932 & 137.674 & 0.0264818\\
4 & -0.016764 & -0.0193729 & 161.317 & 0.072193\\
5 & -0.0119611 & -0.0125862 & 86.4652 & 0.0254668\\
\end{tabular}
\end{ruledtabular}
\bigskip\bigskip
\end{table}

\begin{figure}[t]
\begin{center}
\bigskip\bigskip\bigskip
\includegraphics[width=2.5in]{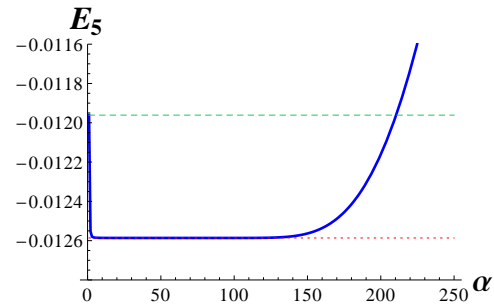}
\caption{(color online) Approximate energy of the fifth state of the "dipole" potential, as a function of the variational parameter $\alpha$, for $n_{max}=60$.
The dotted line corresponds to the minimum value, while the dashed line corresponds to the value for $\alpha =1$, as in ref.~\cite{Dorsey10}.}
\label{Fig_4}
\end{center}
\end{figure}

\begin{figure}[t]
\begin{center}
\bigskip\bigskip\bigskip
\includegraphics[width=3.in]{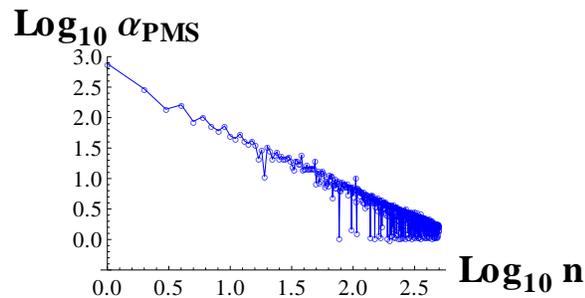}
\caption{(color online) Optimal value of $\alpha$ as a function of $n$.}
\label{Fig_5}
\end{center}
\end{figure}

\begin{figure}[t]
\begin{center}
\bigskip\bigskip\bigskip
\includegraphics[width=2.5in]{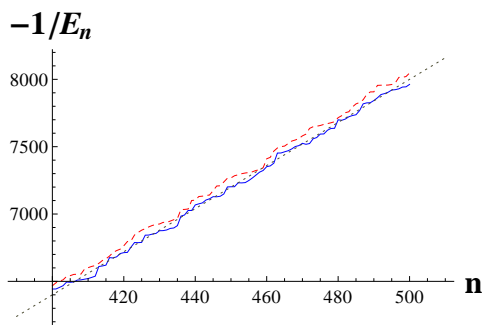}
\caption{(color online) $-1/E_n$ for the excited states between $n=400$ and $n=500$. The solid line is the result obtained with the PMS, while the dashed line is the
result obtained using $\alpha=1$. The dotted line is the asymptotic law of eq. (10) of ref.\cite{Dorsey10}. }
\label{Fig_6}
\end{center}
\end{figure}

In conclusion, we believe that we have presented numerical results which complement and extend the results of ref.~\cite{Dorsey10}; 
we have shown that the application of the Rayleigh-Ritz method to this problem provides very accurate results, both exploiting the Shanks 
transformation and performing the calculation on a selected portion of the Hilbert space. 
In this way we have calculated the ground state energy corresponding
to $n_{max}=150$, working with a matrix of size $885 \times 885$, instead of the full matrix $22500 \times 22500$, obtaining our most precise bound,
$E_1 \leq -0.1291697$. Using the numerical results for the excited states, calculated using the PMS, we have reproduced with high accuracy the
expected asymptotic behavior of eq.(10) of ref.~\cite{Dorsey10}.

\acknowledgments
The author acknowledges the support of SNI-Conacyt.

\end{document}